\documentclass[aps,preprint,nofootinbib,floatfix]{revtex4}
\usepackage{color}

\pdfoutput=1
\usepackage{graphicx}
\usepackage[latin1]{inputenc}
\usepackage{amsmath,amssymb}
\usepackage{hyperref}
\usepackage{epstopdf}
\usepackage{slashed}
\usepackage{cancel}
\usepackage{soul}

\newcommand{\AddrDortmund}{
Fakult\"at f\"ur Physik, Technische Universit\"at Dortmund,
44221 Dortmund, Germany
}

\newcommand{\AddrSDU}{
CP$^{3}$-Origins, University of Southern Denmark, Campusvej 55, DK-5230 Odense M, Denmark
}

\numberwithin{equation}{section}

 \def\be   {\begin{equation}}   \def\ee   {\end{equation}}
       \def\ea   {\end{array}}
 \def\bea  {\begin{eqnarray}}   \def\eea  {\end{eqnarray}}
 \def\bean {\begin{eqnarray*}}  \def\eean {\end{eqnarray*}}

  \def\la   {\lambda}

     \def\De   {\Delta}

 \def\nn{\nonumber}
 \def\lee { \left( }
\def\rii { \right) }
\def\lan   {\langle}
\def\ran   {\rangle}

\def\tol {\leftrightarrow}

\begin{document}

{\small
\begin{flushright}
DO-TH 18/12 \\
CP3-Origins-2018-022 DNRF90
\end{flushright} }

\title{Scalar Dark Matter, GUT baryogenesis and Radiative neutrino mass }

\author{Wei-Chih Huang}\email{huang@cp3.sdu.dk}
\affiliation{\AddrSDU,\AddrDortmund}
\author{Heinrich P\"as\email}\email{heinrich.paes@tu-dortmund.de}
\affiliation{\AddrDortmund}
\author{Sinan Zei{\ss}ner}\email{sinan.zeissner@tu-dortmund.de}
\affiliation{\AddrDortmund}

\begin{abstract}
We investigate an interesting correlation among dark matter phenomenology, neutrino mass generation and GUT baryogenesis,
based on the scotogenic model. The model contains additional
right-handed neutrinos $N$ and a second Higgs doublet $\Phi$, both of which are odd under an imposed $Z_2$ symmetry.
The neutral component of $\Phi$, i.e. the lightest of the $Z_2$-odd particles, is the dark matter candidate.
Due to a Yukawa coupling involving $\Phi$, $N$ and the Standard Model leptons, the lepton asymmetry is converted into the dark matter asymmetry so that a non-vanishing $B-L$ asymmetry can arise from  $(B-L)$-conserving GUT baryogenesis, leading to a nonzero baryon asymmetry after the sphalerons decouple. On the other hand, $\Phi$ can also generate neutrino masses radiatively.
In other words, the existence of $\Phi$ as the dark matter candidate resuscitates
GUT baryogenesis and realizes neutrino masses.   \\

\end{abstract}

\maketitle

\section{Introduction}\label{sec:Introduction}
The origin of the observed baryon asymmetry can not be
accounted for within the Standard Model~(SM) and is one of the unresolved issues in particle physics and cosmology.
The simplest Grand Unified Theory~(GUT) based on the SU(5) model, proposed by Georgi and Glashow in 1974~\cite{Georgi:1974sy}, features
 leptoquark gauge bosons which do mediate baryon number violating processes, leading to proton decay.    
The model, however, conserves the difference between the baryon and lepton number $B-L$. In other words,
any generation of a baryon asymmetry from heavy gauge or Higgs boson decays, as discussed in Refs.~\cite{Yoshimura:1978ex,Toussaint:1978br,Weinberg:1979bt,Barr:1979ye},
comes with an equal amount of  lepton asymmetry.  
These baryon and lepton asymmetries, however, will be washed out completely by non-perturbative sphaleron processes~\cite{Klinkhamer:1984di, Arnold:1987mh,  Arnold:1987zg}, which come into thermal equilibrium when the temperature of the universe drops roughly below $10^{12}$ GeV.
The $B-L$ symmetry conservation also exists in larger symmetry groups, such as $SO(10)$, where the abelian $U(1)_{B-L}$ is a subgroup.
Therefore, as long as $U(1)_{B-L}$ is not broken when a baryon asymmetry is created, i.e., initially $B+L \neq 0$
but $B-L=0$,  such a baryon asymmetry will not survive the sphaleron processes.

In principle, there are at least two ways to revive GUT baryogenesis. First, nonzero $B-L$ can still be realized in certain matter representations under $SO(10)$ or larger groups
as demonstrated, for instance, in Refs~\cite{Coughlan:1985hh,Babu:1992ia,Garbrecht:2005rr,Achiman:2007qz,Babu:2012vb,Babu:2012vc}.
Second, Fukugita and Yanagida~\cite{Fukugita:2002hu}~(and a recent update, Ref.~\cite{Huang:2016wwj})
have proposed to include right-handed neutrinos to resuscitate GUT baryogenesis, where the right-handed neutrino $N$ can be embedded into $SU(5)$
as a singlet or into the $\bold{16}$ of $SO(10)$.
A Majorana mass of $N$, which can arise from the spontaneous symmetry breaking of $L$ via the vacuum expectation value of a scalar or can simply be imposed by hand, explicitly  violates the original $B-L$ symmetry.

\begin{figure}[!htb!]
\centering
\includegraphics[width=0.45\textwidth]{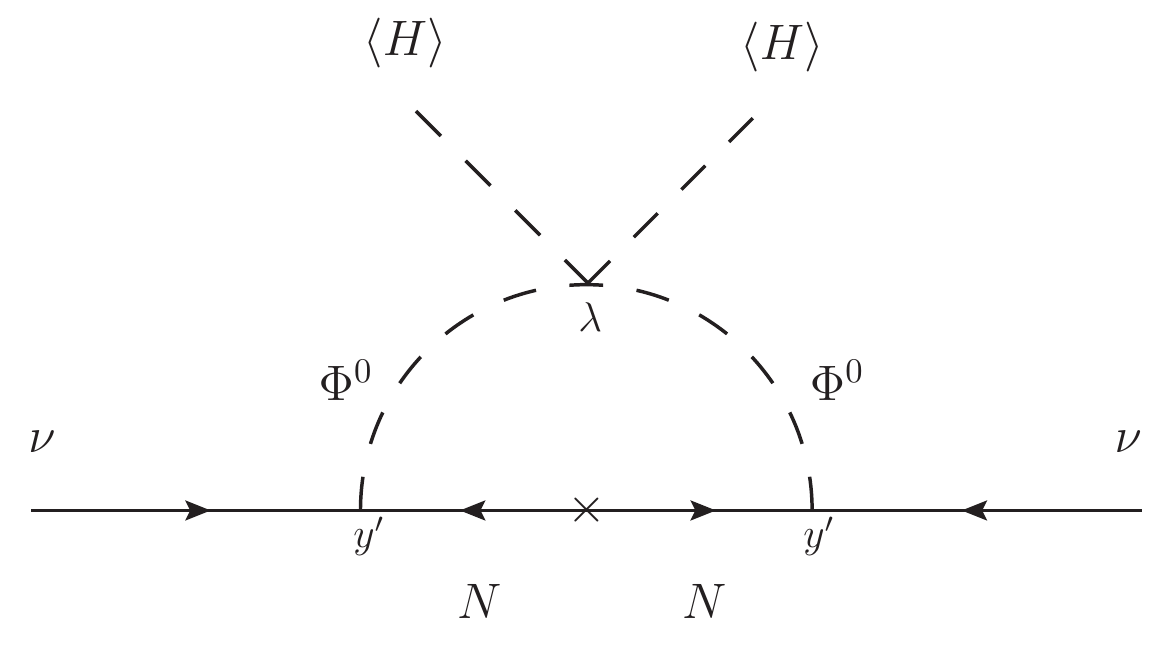} \\
\caption{Neutrino mass generation in the scotogenic model. Here, $y^\prime$ is the Yukawa coupling and $\la$ denotes the quartic  coupling between the Higgs~($H$) and the second
doublet~($\Phi$).}
\label{fig:nu_mass}
\end{figure}

In this paper, we revisit and extend the idea of Fukugita and Yanagida~\cite{Fukugita:2002hu} in the context of
the scotogenic model~\cite{Ma:2006km}. In this model, a second scalar $SU(2)_L$ doublet $\Phi$
is introduced which radiatively generates neutrino masses as shown in Fig.~\ref{fig:nu_mass}. At the same time, the neutral component
of the doublet is a suitable dark matter~(DM) candidate because of
an imposed $Z_2$ symmetry.
\\ 

\begin{figure}[!htb!]
\centering
\includegraphics[width=0.45\textwidth]{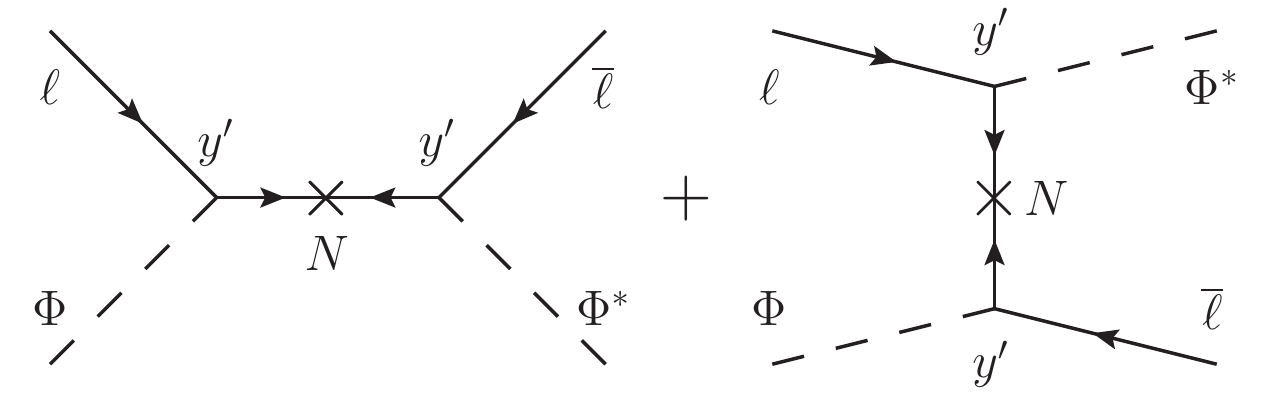}
\includegraphics[width=0.45\textwidth]{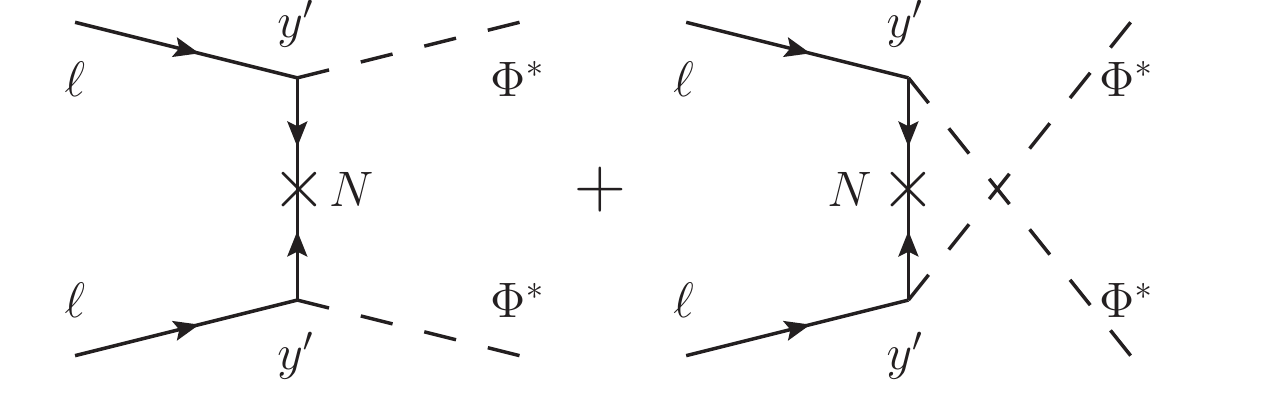}
\caption{ $N$-mediated lepton number violating processes which transfer a $L$ asymmetry into that of DM~($\Phi$). }
\label{fig:N_washout}
\end{figure}

In that both $\Phi$ and right-handed neutrinos $N$ are $Z_2$-odd,
the type-I seesaw Yukawa coupling $y \bar{\ell} H N$ is forbidden~($\ell$: SM lepton doublet) but a new Yukawa coupling $y^\prime \bar{\ell} \Phi N$ is allowed, which induces a washout of lepton number.
As illustrated in Fig.~\ref{fig:N_washout}, the change of the lepton number
is accompanied by a change of the $\Phi$ number,
$\De L = \De \Phi$. In other words, the $L$ asymmetry is transferred into a DM asymmetry.
Moreover, part of the DM asymmetry further shifts to an asymmetry of the Higgs doublet (note that both doublets are equally charged under the SM gauge groups) because of the $\Phi-H$ interactions:
$\Phi^{*} H \tol \Phi H^{*} $ and $\Phi^{(*)}\Phi^{(*)} \tol H^{(*)} H^{(*)} $.
In this scenario, the $L$ asymmetry can be maximally reduced down to one third of the initial value~(instead of one-half in the case without $\Phi$ where only the $y \bar{\ell} H N$ coupling exists~\cite{Fukugita:2002hu,Huang:2016wwj})
since $\Phi$ and $H$ share the asymmetry. That is, the resulting final $B-L$ asymmetry can be maximally one third of the initial $B+L$ asymmetry generated by GUT baryogenesis.
Taking into account the top~(bottom) Yukawa coupling, which is in thermal equilibrium for temperatures $T \lesssim 10^{16}~(10^{12})$ GeV,
the $H$ asymmetry will be transferred into quarks, leading to a larger lepton number washout.
See Ref.~\cite{Huang:2016wwj} for more details.

Note that there exist many models that realize radiative neutrino masses and DM with discrete symmetries.
It has been shown~\cite{Ma:2015xla} in some of these models including the scotogenic model, the dark parity, used to protect the DM stability,
is actually related to lepton number $L$ 
as $(-1)^{L+2j}$, where $j$ is the particle spin. In other words, $L$ and the DM parity are correlated and that is the reason why the lepton asymmetry is
converted into the DM asymmetry in this work.
\begin{figure}[!htb!]
\centering
\includegraphics[width=0.9\textwidth]{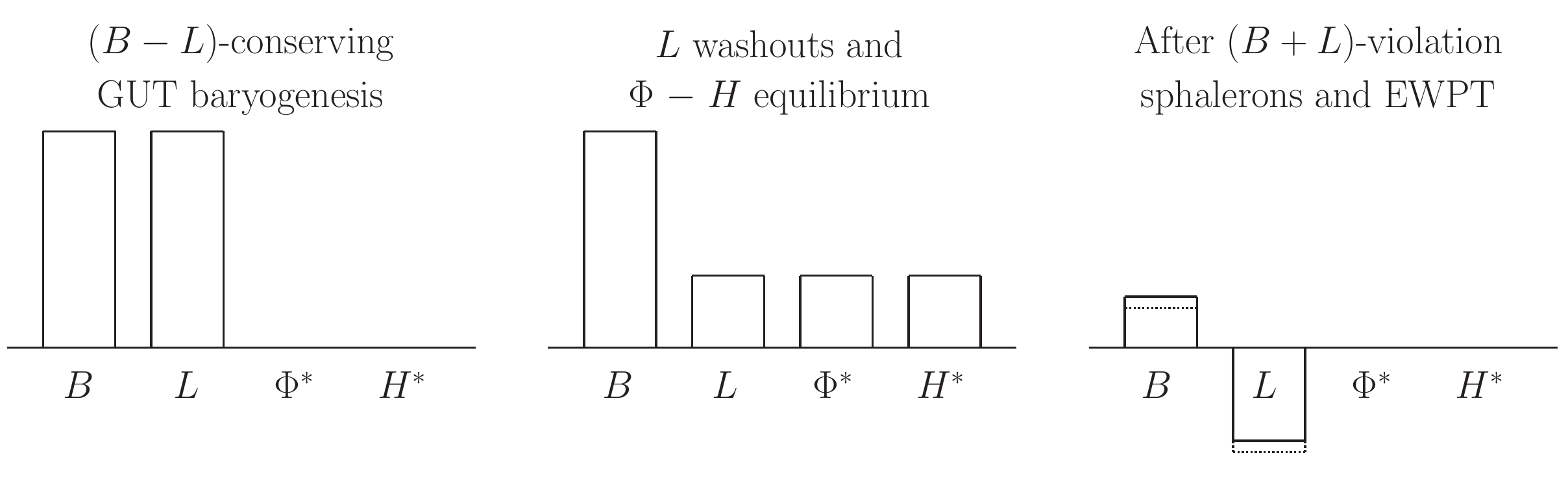} \\
\caption{ Pictorial illustration of  asymmetry conversion in the presence of DM $\Phi$ and the right-handed neutrino $N$. The $L$ asymmetry generated from GUT baryogenesis is converted into a DM~($\Phi$) asymmetry, and then is also shared by $H$ due to
$\Phi-H$ equilibrium.
As a result, the maximal $B-L$ asymmetry is one third of
the initial $B+L$ asymmetry from GUT baryogenesis as indicated in the middle panel.
If DM decouples before the EWPT, the asymmetry will be transferred back to the SM sector, increasing the final $B$ asymmetry as displayed in the right panel, where the solid~(dashed) line corresponds
to DM freeze-out before~{(after)} the EWPT. See the text for more details.}
\label{fig:Asym_DM}
\end{figure}

\begin{figure}[!htb!]
\centering
\includegraphics[width=0.25\textwidth]{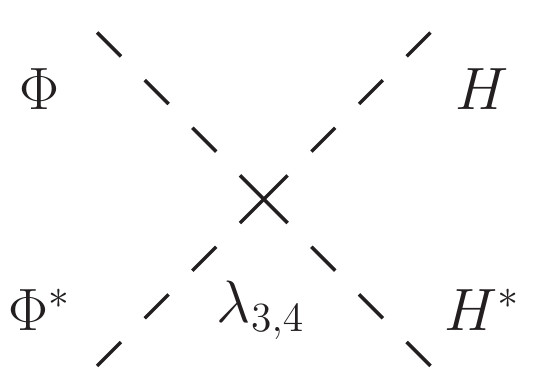}
\includegraphics[width=0.25\textwidth]{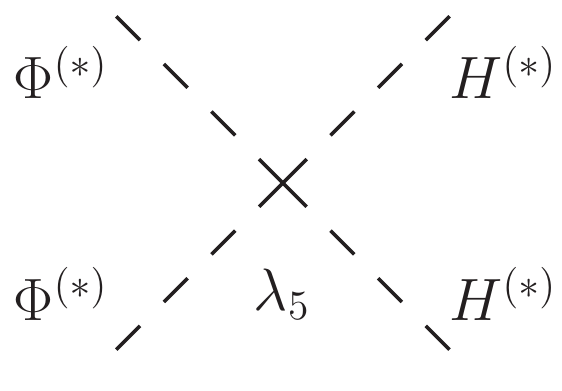}
\caption{ Main DM annihilation processes which determine the DM density. The $\lambda$'s denote the Higgs-DM quartic couplings,
defined in Eq.~\eqref{eq:IH_pot}.}
\label{fig:DM_fre}
\end{figure}

If DM decouples from the thermal bath before the electroweak phase transition~(EWPT), the DM asymmetry will be transformed back to
$H$ via the process $\Phi H^* \tol \Phi^* H$, which has only a single power of Boltzmann suppression and is very efficient compared to the doubly Boltzmann-suppressed annihilation channels     
$\Phi \Phi^{*} \tol H H^*$ and $\Phi^{(*)}\Phi^{(*)} \tol H^{(*)} H^{(*)} $. This conversion will slightly increase the final baryon asymmetry because the Yukawa couplings and the sphalerons will redistribute the asymmetries among quarks, leptons and Higgs bosons.
Note that after the EWPT, the asymmetry for the real part of the neutral component $H^0$ will vanish because of
the Higgs vacuum expectation value~\cite{Harvey:1990qw},
whereas the remaining degrees of freedom of $H$ will become the longitudinal component of $W^{\pm}$ and $Z$.
Similarly, the $\Phi^0$~(neutral components of $\Phi$) asymmetry will also vanish after the EWPT due to the efficient $\Phi-H$
interactions\footnote{The interaction $\Phi^0 h \leftrightarrow \Phi^{0*} h$~($h$: SM Higgs boson after the EWPT) will erase the $\Phi^{0(*)}$ asymmetry.} 
while the $\Phi^\pm$~(charged components) asymmetry will move to $W^\pm$. The final $B$ and $L$ asymmetries will stay unchanged since the sphalerons become ineffective after the EWPT.
Fig.~\ref{fig:Asym_DM} elucidates the asymmetry transformation as a function of time. On the other hand, the DM relic abundance
is mainly determined by the Higgs-DM couplings for TeV DM (such that DM freezes out prior to the EWPT)
as shown in Fig.~\ref{fig:DM_fre}. 

Note that the scotogenic model alone can generate the baryon asymmetry, apart from realizing neutrino masses  and accommodating DM candidates, via heavy neutrino decays  as first pointed out by Ref.~\cite{Ma:2006fn}  and followed by more detailed studies~\cite{Kashiwase:2012xd,Kashiwase:2013uy,Racker:2013lua,Clarke:2015hta}. 
The subject has been further developed recently -- Ref.~\cite{Hugle:2018qbw} which attains low-scale leptogenesis without any degeneracy in the right-handed neutrino mass spectrum and Ref.~\cite{Baumholzer:2018sfb} which features the KeV right-handed neutrino as a DM candidate.
In this work, we instead focus on the $L$ washout effects induced by the heavy neutrinos, and assume the new Yukawa coupling is CP-conserving. In other words, the right-handed neutrino decays  equally into leptons and antileptons\footnote{Note that, while this paper focuses on the asymmetry transfer between DM and leptons without considering the decay contribution, in the presence of CP violation in the Yukawa couplings, the asymmetry from N decays could be sizable. As pointed out in Ref.~\cite{Sierra:2013kba}, tree-level $N$-mediated washout processes are out of equilibrium during the time of $N$ decays if $m_N \gtrsim 10^7$ GeV such that the $L$ asymmetry from the decays can survive from washouts and account for the observed baryon asymmetry.
The region of interest in this work, $m_N \gtrsim 10^{10}$ GeV, falls into this region and hence the decay effect could be important, depending
on the values of the Yukawa couplings and the size of the CP phase(s).}.

This paper is organized as follows. In Section~\ref{sec:model}, we briefly review the scotogenic model and then develop the formalism for lepton number washout based on Boltzmann equations in Section~\ref{sec:washouts}. In Section~\ref{section:asym_transfer}, we explain how asymmetries are transferred between the DM and SM sectors and  present our numerical results of the Boltzmann equations. The relic abundance is calculated in Section~\ref{sec:DM_rel} where the DM direct search bounds from the XENON1T experiment are also taken into account.
Finally, we conclude in Section~\ref{sec:conclusion}.

\section{Scotogenic model}\label{sec:model}

The scotogenic model has been proposed by E. Ma~\cite{Ma:2006km}, where the neutrino mass is loop-induced by a second
$SU(2)_L$ doublet scalar $\Phi$ and the right-handed neutrinos $N$, both of which are odd under an imposed $Z_2$ symmetry. Thus, Yukawa couplings in the type-I
seesaw, $y_{i j} \bar{\ell}_i H^* N_j$ are forbidden and replaced by $y^\prime_{i j} \bar{\ell}_i \Phi^* N_j$.
In principle both the $Z_2$-odd $N$ and the neutral component of the $\Phi$ doublet could be the DM candidates.
However, the mass of $N$ being of interest for this work is above $10^{10}$ GeV, that is too heavy to thermally generate the correct
relic density~\cite{Griest:1989wd}.
In the framework of $SU(5)$,  $\Phi$ can be embedded into the representation of
$\bold{5}$, while $N$ can be a singlet. We here simply assume that other particles,
which are embedded in the same representation of $SU(5)$~(or larger symmetry groups) as SM particles or $\Phi$,
are heavier than the scale of interest. Thus, 
only the SM particles, $\Phi$ and $N$ are taken into account in the analysis.

In addition to the SM interactions, the Lagrangian reads
\begin{align}
\mathcal{L} \supset y^\prime_{i j} \bar{ \ell }_i \Phi^* N_j + \frac{M_{N_k}}{2} \overline{N^c_k} N_k + V \lee \Phi, H \rii,    
\end{align}
with
\begin{align}
V\lee H, \Phi \rii =  \mu^2_1 \vert H \vert^2 + \mu^2_2 \vert \Phi \vert^2 + \la_1 \vert H \vert^4 + \la_2 \vert \Phi \vert^4
+ \la_3 \vert H \vert^2 \vert \Phi \vert^2 + \la_4 \vert H^* \Phi \vert^2 + \frac{\la_5}{2} \lee \lee H^* \Phi\rii^2 + \text{h.c.}  \rii , 
\label{eq:IH_pot} 
\end{align}
which is just the scalar potential of the inert Higgs Doublet model~\cite{Deshpande:1977rw}.
The radiative neutrino mass matrix induced by loops of $\Phi$ and $N$ is~\cite{Ma:2006km}\footnote{Note that there is a factor of $1/2$ missing in Ref.~\cite{Ma:2006km}; see, e.g., version 1 of Ref.~\cite{Merle:2015gea} or Appendix C of Ref.~\cite{Vicente:2015zba}.}
\begin{align}
\lee m_\nu \rii_{ij} = \sum_k \frac{ \lee y^\prime_{ik} y^\prime_{jk} \rii^* M_{N_k} } { 32 \pi^2 }
\lee \frac{m^2_R}{ m^2_R - M^2_{N_k}} \log\frac{m^2_R}{M^2_{N_k}} - \frac{m^2_I}{ m^2_I - M^2_{N_k}} \log\frac{m^2_I}{M^2_{N_k}}  \rii ,
\end{align}
where
\begin{align}
m^2_R&= \mu^2_2 + \frac{1}{2} \lee  \la_3 + \la_4 +\la_5 \rii v^2 , \nn \\
m^2_I&= \mu^2_2 + \frac{1}{2}\lee  \la_3 + \la_4  - \la_5 \rii v^2 ,
\label{eq:two_neu}
\end{align}
with $v=246$ GeV being the Higgs vacuum expectation value.
Note that in order to obtain a non-vanishing neutrino mass, one must have $m_R \neq m_I$, i.e., $\lambda_5 \neq 0$.
We here are interested in the region  of $M_N \gtrsim 10^{10}$ GeV, $m_R \sim m_I\sim$ TeV and
$\vert m_R - m_I\vert \ll m_I \sim m_R $.
In this case, the neutrino mass matrix becomes
\begin{align}
\lee m_\nu \rii_{ij} =  \frac{ \la_5 v^2 } { 32 \pi^2 } \sum_k
\frac{ \lee y^\prime_{ik} y^\prime_{jk} \rii^* } { M_k } 
\lee \log\lee \frac{M^2_{N_k}}{m^2_0}  \rii  -1  \rii ,
\label{eq:loop_mnu}
\end{align}
where $m_0= \frac{m_R + m_I}{2}$. To reproduce the observed neutrino mass squared difference responsible for atmospheric neutrino oscillations, the heaviest neutrino must be heavier than $0.05$ eV or so, which corresponds to $\la_5 \sim 6 \times 10^{-3} $ for $M \sim 10^{12}$ GeV and $m_0 \sim $ TeV, given
$y^\prime$ of $\mathcal{O}(1)$.

\section{Washout Formalism} \label{sec:washouts}

Due to the Hubble expansion, a convenient quantity to describe the particle number density is $Y \equiv n/s$, which is the
number density normalized to the entropy density $s$, i.e., the number per co-moving volume. The density $Y$ is conserved
in the absence of particle creation or annihilation.
The Boltzmann equation of a particle $\ell$
for an interaction $\ell a_1 \cdots a_n \leftrightarrow f_1 \cdots f_m$ is,  
\begin{align}
\label{eq:BoltzmannY}
	z H s \frac{d Y_{\ell}}{dz} = -\sum_{a_i,f_j} [ \ell a_1 \cdots a_n \leftrightarrow f_1 \cdots f_m ]\,,
\end{align}
where $H$ is the Hubble parameter, $z = M_{N}/T$, and 
\begin{align}
	[  \ell a_1 \cdots a_n \leftrightarrow f_1 \cdots f_m] 
	&=
	\frac{n_\ell n_{a_1} \cdots n_{a_n}}{n_\ell^{\rm eq} n_{a_1}^{\rm eq}\cdots n_{a_n}^{\rm eq}} 
	\gamma^\text{eq}(  \ell a_1 \cdots a_n \leftrightarrow f_1 \cdots f_m)                   \nonumber\\
	&- \frac{n_{f_1} \cdots n_{f_m}}{n_{f_1}^{\rm eq} \cdots n_{f_m}^{\rm eq}}
      \gamma^\text{eq}\left( f_1 \cdots f_m \leftrightarrow \ell a_1 \cdots a_n \right).
\end{align}
The thermal rate $\gamma^{\text{eq}}$ is defined as
\begin{align}
	\gamma^{\text{eq}}(\ell a_1 \cdots a_n \to f_1 \cdots f_m) 
	&=    \Big[ \int \frac{\mathrm{d}^3 p_{\ell}}{2 E_{\ell} (2\pi)^3} e^{-\frac{E_{\ell}}{T}} \Big] \prod\limits_{a_i} 
	\Big[ \int \frac{\mathrm{d}^3 p_{a_i}}{2 E_{a_i} (2\pi)^3} e^{-\frac{E_{a_i}}{T}} \Big] \nonumber\\ \times \prod\limits_{f_j} 
	\Big[ &\int \frac{\mathrm{d}^3 p_{f_j}}{2 E_{f_j} (2\pi)^3} \Big] 
        \times (2\pi)^4\delta^4 \Big( p_{\ell} + \sum_{i=1}^n p_{a_i} - \sum_{j=1}^m p_{f_j} \Big) |M|^2,
\label{eq:ga_def}               
\end{align}
where $|M|^2$ is the squared amplitude summing over initial and final spins.

To simplify the analysis, we consider a $1+1$ scenario, i.e., one generation of the SM leptons and one right-handed
neutrino\footnote{For simplicity, we stick to the cases where the initial lepton asymmetry is stored in the lepton doublet.}.
Moreover, we assume that the scale of GUT baryogenesis is slightly below the right-handed neutrino mass
to avoid complications from finite-temperature effects (if, for example, the decay
$N \to H L$ would be kinematically forbidden, the first processes in Fig.~\ref{fig:N_washout} would not have resonance anymore, reducing the $L$ washout effect) due to thermal masses when $T  \gtrsim m_{N}$~\cite{Giudice:2003jh}.

For the $L$ washout computation, we include both $\Delta L=1$ and $\Delta L=2$ interactions. Following the notation of Ref.~\cite{Giudice:2003jh},
the $\De L =2$ washout processes include $\ell \Phi \tol \bar{\ell} \Phi^*$~(with thermal rate $\gamma_{Ns}$)
and $\ell \ell  \tol \Phi^* \Phi^*$ $(\gamma_{Nt})$ as displayed in Fig.~\ref{fig:N_washout}. The relevant $\De L = 1$ washout processes are
$\ell \Phi \tol N$ $\left(\gamma_D \right)$, $\ell N \tol \Phi^* A$ $(\gamma_{As})$, $\ell \Phi \tol N A$ $(\gamma_{At_1})$
and $\ell A \tol N \Phi^*$ $(\gamma_{At_2})$.
We refer readers to our previous work~\cite{Huang:2016wwj} and references therein for more details. Note that the previous work is based on the type-I seesaw mechanism while in this work, it is another Yukawa coupling $y^\prime \ell \Phi^* N$ that is responsible for the washout processes.
The formalism of washout computation is, however, similar for the two cases.

The resulting Boltzmann equations including the lepton washout and sphalerons~\cite{Moore:2000ara,D'Onofrio:2014kta}
processes read 

\begin{align}
zHs \frac{d Y_{B- L}}{dz} = & 2 \left( 2 \gamma_{Ns} + 4 \gamma_{Nt} + \gamma_{As} \frac{Y_N}{Y_N^{\text{eq}}} + \gamma_{At_1} + \gamma_{At_2} \right) \frac{Y_{B+L}-Y_{B-L}}{2Y_L^{\text{eq}}} \nonumber \\
 &- 2 b_\Phi \bigg[ \gamma_{Ns} + 4 \gamma_{Nt} + \gamma_{As}+ \gamma_{At_1} + \gamma_{At_2}  \frac{Y_N}{Y_N^{\text{eq}}}  \bigg] \frac{Y_{\Phi'}}{Y_{\Phi}^{\text{eq}}} \ ,\\
zHs \frac{d Y_{B+L}}{dz} = &- 2 \left( 2 \gamma_{Ns} + 4 \gamma_{Nt} + \gamma_{As} \frac{Y_N}{Y_N^{\text{eq}}} + \gamma_{At_1} + \gamma_{At_2} \right) \frac{Y_{B+L}-Y_{B-L}}{2Y_L^{\text{eq}}} \nonumber \\
 &+ 2 b_\Phi \bigg[ 2 \gamma_{Ns} + 4 \gamma_{Nt} + \gamma_{As}+ \gamma_{At_1} + \gamma_{At_2}  \frac{Y_N}{Y_N^{\text{eq}}}  \bigg] \frac{Y_{\Phi'}}{Y_{\Phi}^{\text{eq}}} 
 + \frac{351}{2} \alpha_W^5 \frac{M_N s}{z} Y_{B+L} \ ,\\
  zHs \frac{d Y_{N}}{dz}  =  &- \left( \gamma_D + 4\gamma_{As} + 4\gamma_{At_1} + 4\gamma_{At_2} \right) \left(\frac{Y_{N}}{Y_N^{eq}}-1\right) \ , \\
 \frac{d Y_{\Phi'}}{dz} \equiv &  \frac{d Y_{B-L}}{dz} \ ,
\end{align}
where $Y_{L(B)} \equiv Y_{\text{lepton~(baryon)}} - Y_{\text{anti-lepton~(anti-baryon)}}$ and $Y^{\text{eq}}$
is the equilibrium density of the corresponding (anti-)particle.
The impact of the $t$- and $b$-Yukawa couplings on the washout processes can be characterized by the factor $b_{\Phi}$~\cite{Huang:2016wwj}:
\begin{align}
b_{\Phi} = \left\lbrace \begin{matrix}
&\frac{1}{3} &\;\;  \;\; 10^{12} \lesssim T  \lesssim 10^{16}\text{ GeV}\\
&\frac{1}{5} &\;\; \;\; T \lesssim 10^{12}\text{ GeV}
\end{matrix} \right. \;\; .
\end{align}
Moreover, the chemical equilibrium for $\Phi^{(*)} \Phi^{(*)} \tol H^{(*)} H^{(*)}$ is reached if
\begin{align}
\frac{\lambda_5^2 T}{8 \pi} \gtrsim \frac{T^2}{m_{\text{Pl}}} ,
\end{align}
with $m_{\text{Pl}}= 1.22 \times 10^{19}$ GeV.
The $\Phi-H$ chemical equilibrium is always fulfilled for values of $\la_5$ of interest.

The final $B$ and $L$ asymmetries as functions of the final $B-L$ asymmetry are
\begin{align}
Y^{\text{final}}_B= c_s Y^{\text{final}}_{B - L} \;\; , \;\;  Y^{\text{final}}_L= (c_s -1) Y^{\text{final}}_{B - L} \; .
\end{align}
For non-supersymmetric models the sphaleron conversion factor is $c_s= 28 / 79$~\cite{Khlebnikov:1988sr, Harvey:1990qw} if DM decouples before the EWPT.
On the other hand, if DM freezes out after the EWPT, the $\Phi^0$~($\Phi^\pm$) asymmetry will just vanish~(transfer into $W^\pm$),
and has no influence on the final $B$ and $L$ asymmetries
as explained above. In this case, one has $c_s=8/23$ as we shall see below.

\section{Asymmetry transfer between DM and SM sectors}
\label{section:asym_transfer}

We now are in the position to explain how the washout processes can create a nonzero $B-L$ asymmetry
and how asymmetries are transferred among different particles.

For temperatures above $10^{12}$ GeV, the $(B+L)$-violating\footnote{In the following, we will use the shorthand notations
$\cancel{B+L}$, $\cancel{B}$ and $\cancel{L}$ for $(B+L)$-, $B$- and $L$-violating, respectively.}
sphalerons are not in thermal equilibrium and part of the lepton asymmetry is moved to DM due to the washout processes induced by the Yukawa coupling $y^\prime \bar{\ell} \Phi N$.
For both of the $\Delta L=1$ and $\Delta L=2$ interactions, the change in the lepton number comes with
an equal amount of the DM number change. The partial asymmetry of $\Phi$ is further converted into $H$ through the interaction, $\frac{\la_5}{2} \lee \lee H^* \Phi\rii^2 + \text{h.c.}  \rii$.  
That is, after $L$ washout one obtains a nonzero $B-L$ asymmetry: $\Delta(B-L)   = - (\Delta \Phi + \Delta H)$.

\begin{figure}[!htb!]
\centering
\includegraphics[width=0.4\textwidth]{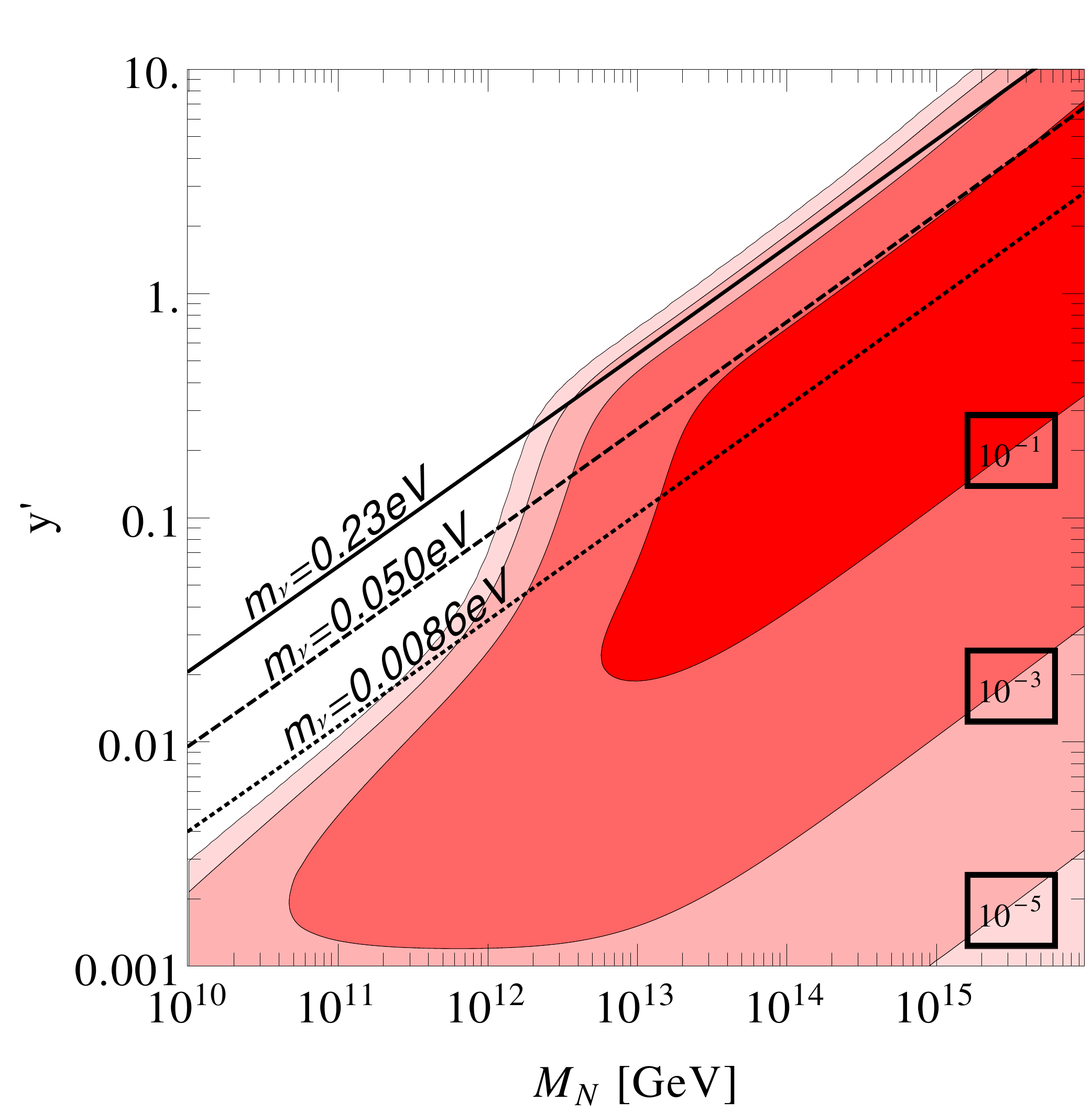}
\includegraphics[width=0.4\textwidth]{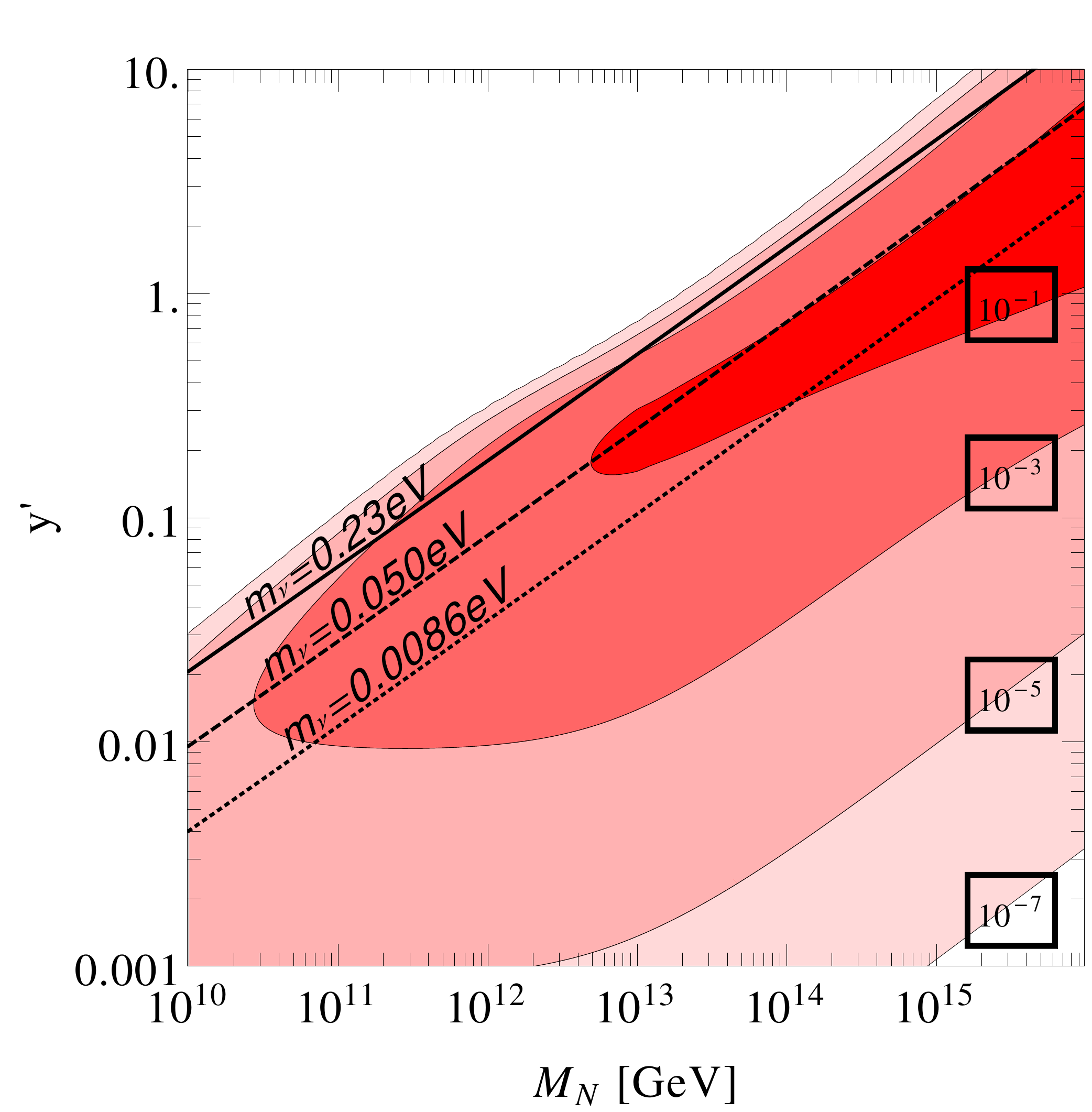}
\caption{ Contour plots of $Y^{\text{final}}_{B-L}/Y^{\text{initial}}_{B+L}$ with $\la_5$ set to unity and $m_0=5$ TeV. The left~(right)
panel refers to the case of the initial $B+L$ asymmetry being created at the scale of $M_N/3$~($M_N/10$). 
The black solid, dashed and dotted lines present the neutrino mass of $0.23$ eV, $\sqrt{\De m^2_{atm}} \simeq 0.05$ eV and $\sqrt{\De m^2_{sol}} \simeq 8.6 \times 10^{-3}$ eV. See the text for more details.}
\label{fig:Boltz_BL_DM}
\end{figure}

For the washout calculation, $m_0= 5$ TeV and $\la_5 = 1$ are assumed which guarantee that the $\Phi-H$ interactions are always in chemical equilibrium during the period of washout, i.e., $\Delta(B-L)   =- 2 \Delta \Phi$.
The numerical results are presented in Fig.~\ref{fig:Boltz_BL_DM} with the initial $B+L$ asymmetry from GUT baryogenesis injected at the scale of $M_N/3$~(left panel) and $M_N/10$~(right panel). 
The contours represent the ratio of the final $B-L$ to the initial $B+L$ asymmetry,
i.e., $Y^{\text{final}}_{B-L}/Y^{\text{initial}}_{B+L}$.
A smaller $B+L$ injection scale implies a shorter $L$ washout period before the sphalerons kick in, and hence requires a larger Yukawa
coupling~(a higher washout rate) to compensate. As a result, the $Y^{\text{final}}_{B-L}/Y^{\text{initial}}_{B+L}$ contours move upward in the right panel
when compared to the left one.

Maximal $L$ washout~(the maximal final $B-L$ asymmetry) denoted by dark red areas arises from the case in which the $\slashed{L}$ processes are very efficient before the sphalerons come into play but become ineffective when the sphalerons are in thermal equilibrium.
Two minimal $B-L$ asymmetry scenarios~(white areas) correspond to situations where first $L$ washouts have never been fast enough
before the sphalerons destroy most of the initial $B+L$ asymmetry, and second both $\slashed{L}$ and $\cancel{B+L}$ processes are present and effective for a long time, leading to vanishing $B$ and $L$ asymmetries\footnote{Again, the detailed analysis can be found in our previous work, Ref.~\cite{Huang:2016wwj}, with the different particle contents
but with a similar washout mechanism.}.
The black solid line indicates the active neutrino mass $m_\nu$ of 0.23 eV, the bound from Planck~\cite{Ade:2015xua} on the sum of the active neutrino masses,
while the black dashed line corresponds to
$m_\nu = \sqrt{\De m^2_{atm}} \simeq 0.05$ eV and the black dotted line for $m_\nu = \sqrt{\De m^2_{sol}} \simeq 8.6 \times 10^{-3}$ eV.
If $\la_5$ is increased~(decreased),
according to Eq.~\eqref{eq:loop_mnu} the black lines will move downwards~(upwards) accordingly. On the other hand, the active neutrino masses are not very sensitive to the precise value of $m_\Phi$ due to the logarithmic dependence.

In order to obtain $m_{\nu} = 0.05$ eV and $Y^{\text{final}}_{B-L}/Y^{\text{initial}}_{B+L} \gtrsim 10^{-2}$,
$M_N$ has to be roughly above $10^{13}$ GeV with $y^\prime \sim 0.25$.
In our previous work~\cite{Huang:2016wwj} with the type-I seesaw Yukawa coupling $\bar{\ell} H^* N$,
one can achieve larger washout effects with $Y^{\text{final}}_{B-L}/Y^{\text{initial}}_{B+L} \gtrsim 10^{-1}$ and at the same time reproduce $m_{\nu} = 0.05$.
The main difference in the presence of $\Phi$ is that the active neutrino mass is loop-induced and hence a larger Yukawa coupling is needed.
In this case, the washout processes last for a longer time and coexist with the \cancel{B+L} sphalerons, leading to a smaller $B-L$ asymmetry.

Depending on the initial $B+L$ asymmetry, there exist regions of the parameter space capable of reproducing $Y^{\text{final}}_{B-L} \simeq 2.4 \times 10^{-10}$ to account for the observed baryon asymmetry, $Y^{\text{final}}_{B} = 8.7 \times 10^{-11}$~\cite{Ade:2015xua}.   
Assuming that, for example, the initial $B+L$ asymmetry is of order  $10^{-6}$ and the $B+L$ injection scale is $M_N/10$, $M_N$ can be as low as $10^{10}$ GeV to realize both the baryon asymmetry and the neutrino mass. In this case $L$ washouts can still be efficient enough to generate a non-vanishing $B-L$ asymmetry before the sphalerons destroy the entire $B+L$ asymmetry.

When the temperature drops below $10^{12}$ GeV and becomes much smaller than $M_N$, $L$ washouts are ineffective but the sphaleron processes start to destroy the $B+L$ asymmetry.
Later on, SM Yukawa couplings reach equilibrium to rearrange the asymmetry among leptons, quarks and the Higgs boson.
One can repeat the analysis of chemical equilibrium done in Refs.~\cite{Khlebnikov:1988sr, Harvey:1990qw,Buchmuller:2005eh}, including an
extra constraint, $\mu_\Phi=\mu_H$.
To simplify the analysis, we assume universal chemical potentials $\mu_\ell$ and $\mu_{e_R}$
for the three left-handed lepton doublets and three right-handed leptons, respectively,   
and all the Yukawa couplings are in thermal equilibrium. This yields 
 \begin{align} 
 &\mu_q= - \frac{1}{3} \mu_{\ell}, \;\;
 \mu_{u_R}=  \frac{1}{6} \mu_{\ell} , \;\;  \mu_{d_R}=  - \frac{5}{6} \mu_{\ell},  \nonumber \\
 & \mu_{e_R} = \frac{1}{2} \mu_{\ell} , \;\;  \mu_{H}=  \frac{1}{2} \mu_{\ell} , \;\;    \mu_{\Phi}=  \frac{1}{2} \mu_{\ell},  \\
    \label{eq:mu_rel_Phi}
 \end{align}
and thus the final $B$ and $L$ asymmetries are 
\begin{align}
B_{f} &=  \frac{8}{23} \lee B- L\rii , \nn \\
L_{f} &= - \frac{15}{23} \lee B- L\rii ,
\end{align} 
which is different from the case in the absence of $\Phi$ with $B_{f} =  \frac{28}{79} \lee B- L\rii$  and $L_{f} = - \frac{51}{79} \lee B- L\rii$~\cite{Khlebnikov:1988sr, Harvey:1990qw,Buchmuller:2005eh} for the SM.
That is to say, $\Phi$ {\it shares} the asymmetry and slightly reduces the baryon asymmetry for a given $B-L$ asymmetry.

Finally, when the temperature falls below $m_\Phi$, $\Phi$ begins to freeze out of the thermal bath.
The DM relic density will be mainly determined by the quartic couplings $\la_{3,4,5}$  in Eq.~\eqref{eq:IH_pot},
if they are large compared to the gauge couplings and Yukawa couplings.
In other words, the DM particle dominantly annihilates into the Higgs bosons. The interaction terms of $\la_3$ and $\la_4$ apparently will not change any asymmetries in $\Phi$ and $H$ while the $\la_5$ term, corresponding to $\Phi \, H^* \leftrightarrow \Phi^* \, H$ and  $\Phi^{(*)} \, \Phi^{(*)} \leftrightarrow H^{(*)} \,  H^{(*)}$, will shift the asymmetry from $\Phi$ to $H$.
Note that the interaction $\Phi \, H^* \leftrightarrow \Phi^* \, H$ is always much faster than the DM annihilation processes
if $\la_{3,4,5}$ are of the same order. That is because the former interaction is singly Boltzmann-suppressed but the latter ones are doubly suppressed. 
The asymmetry conversion between $\Phi$ and $H$ during freeze-out can be understood in the following simple ways.
Since $\Phi$ and $H$ carry the same $U(1)_Y$ charge, the disappearance of $\De \Phi$ has to be compensated by the equal amount of $\De H$ so that the total $U(1)_Y$ is conserved.

In the case where DM freezes out before the EWPT, the $\Phi$ asymmetry will be transformed into that of $H$ and further into those of the quarks and leptons.
On the other hand, if DM freeze-out takes place after the EWPT,
due to the $\Phi-H$ interactions the $\Phi^0$ asymmetry will simply vanish while the $\Phi^\pm$ asymmetry will
transfer to that of $W^\pm$. Due to the fact that the sphaleron effects are not effective anymore below the EWPT, both the $L$ and $B$ asymmetries are
conserved quantities independent of the $\Phi$ asymmetry. 
The final baryon number will slightly increase by $2\%$ if DM decouples before the EWPT and hence
the asymmetry conversion occurs.

We would like to emphasize that regardless of the decoupling time of DM, the final DM abundance is not related to the baryon asymmetry, even if the initial DM asymmetry is closely connected to the initial $B-L$~(also $B$) asymmetry.
 This is the price we have to pay in order to radiatively generate non-zero active neutrino masses via a non-zero $\la_5$. If $\Phi$ decouples before the EWSB the interaction of $\la_5$ quickly shifts the $\Phi$ asymmetry into that of $H$ as the density of $\Phi$ and $\Phi^*$ decrease during freeze-out. Hence the final density of $\Phi$ is only determined by the annihilation of $\Phi$ and $\Phi^*$, similar to symmetric DM scenarios.
If $\Phi$ decouples after the EWPT the asymmetry stored in $\Phi^0$ and $\Phi^{0*}$ just vanishes due to the $\Phi-H$ interactions as explained above. In addition, a non-zero $\la_5$ will result in a mass splitting between the two neutral components as indicated in Eq.~\eqref{eq:two_neu}.
Thus the lightest neutral component is the DM particle, which is real and is its own antiparticle.
The final DM density will only be determined by the DM annihilation into two Higgs bosons.
Note that a zero $\la_5$ would yield correlation between the final DM abundance and the baryon asymmetry. In this case,
the DM mass has to be around 5 GeV to reproduce the correct relic abundance.
As $\Phi$ is a $SU(2)$ doublet, the $Z$ boson can decay into $\Phi$ $\Phi^*$, increasing the $Z$ decay width.
This, however, will be excluded by the LEP bound. In other words, an asymmetric DM scenario cannot be realized in this framework.

\section{DM relic density and direct detection}
\label{sec:DM_rel}

In this section, we compute the DM relic density and discuss direct search bounds.
The study of DM phenomenology for inert Higgs doublet models after electroweak symmetry breaking has been studied,
for instance, in Refs.~\cite{Arhrib:2013ela,Ilnicka:2015jba,Khan:2015ipa,Belyaev:2016lok,Eiteneuer:2017hoh}, while annihilation cross-sections in an unbroken phase have been computed in Ref.~\cite{Garcia-Cely:2015khw}. We here focus on the scenario in the latter case where TeV $\Phi$ decouples before the EWPT and
the main annihilation channels are $\Phi \Phi^{*} \to H H^{*}$ and $\Phi^{(*)} \Phi^{(*)} \to H^{(*)} H^{*}$ as shown in Fig.~\ref{fig:DM_fre}.
 As we shall see later, to achieve the correct DM density, the DM-Higgs couplings $\la$'s have to be larger than unity and also
 than the gauge and Yukawa couplings. Thus, we neglect the gauge and fermion final states in the computation. 

Since the $\Phi$ asymmetry is basically zero during~(and after) freeze-out, the computation of the DM relic density
$\Omega_{\Phi}+ \Omega_{\Phi^*}$ is essentially the same as in the standard symmetric
DM scenario and can be well approximated~\cite{Jungman:1995df,Bertone:2004pz} by
\begin{align}
\Omega_{\text{DM}}  h^2 \approx 2 \frac{3 \times 10^{-27} \text{cm}^3 \text{sec}^{-1}}{ \lan \sigma v\ran_{\Phi  \Phi^* \to H H^*}} \, .
\end{align}
Here the thermally-averaged annihilation cross-section multiplied by the DM relative velocity is
\begin{align}
\lan \sigma v\ran _{\Phi  \Phi^* \to H H^*} \simeq \frac{ \la^2_3}{ 32 \, \pi \, m^2_\Phi}
\end{align}
where we assume  $\la_3 \gtrsim \la_5$ and $\la_4=0$ for simplicity.
Note that the mass degeneracy among components of $\Phi$ will be lifted after electroweak symmetry breaking.
Heavy components of $\Phi$ will decay into the lightest one but the total relic density stays constant due to the unbroken $Z_2$ symmetry.
In fact, for TeV DM the contribution from the Higgs vacuum expectation value to the DM mass is negligible as can be seen from Eq.~\eqref{eq:two_neu}, i.e., $m_\Phi \simeq m_0$.
\begin{figure}[h]
\centering
\includegraphics[width=10cm]{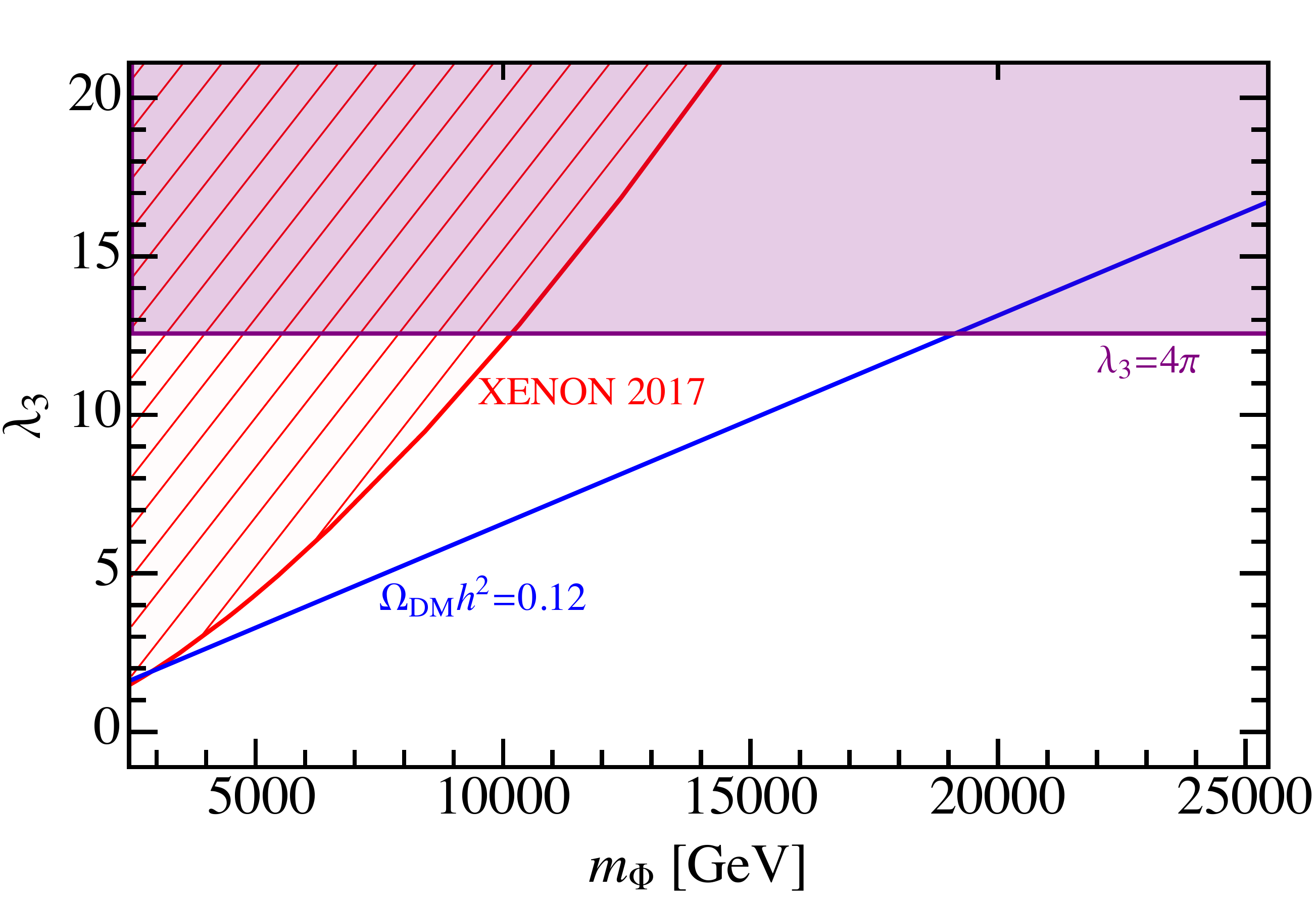}
\caption{Quartic coupling $\la_3$ versus DM mass $m_\Phi$.
The blue line corresponds to the observed relic density while the red dashed area is excluded by the XENON1T
direct search result~\cite{Aprile:2017iyp}. The purple line represents the perturbativity limit $4 \pi$. } \label{fig:la3_mph}
\end{figure}

On the other hand, $\Phi$ can interact with nucleons through the Higgs exchange and null results from DM direct searches put constraints
on the DM-Higgs coupling $\la_3$. Again with the assumption of $\la_3 \gtrsim \la_5$ and $\la_4=0$,
the DM-nucleon spin-independent cross-section is~\cite{Cline:2013gha}
\begin{align}
\sigma_{\text{SI}} = \frac{\la^2_3 f^2_N}{4 \, \pi} \frac{\mu^2 m^2_n}{m^4_h m^2_{\Phi}},
\label{eq:SI_Xsect}
\end{align}
where $f_N=0.3$, $\mu= m_n m_\Phi /(m_n + m_\Phi)$ and $m_n$ is the nucleon mass.

In Fig.~\ref{fig:la3_mph}, we show the direct search bound from the XENON1T result~\cite{Aprile:2017iyp}
 denoted by the red line\footnote{The PandaX-II~\cite{Cui:2017nnn} and LUX~\cite{Akerib:2016vxi} experiments
 yield similar limits, while the latest XENON1T result~\cite{Aprile:2018dbl} only presents the bound for DM below 1 TeV.} and the blue line corresponds to the correct relic density, while the purple line is the perturbativity limit.
 It is clear that XENON1T is unable to probe the large DM mass region
 as the DM-nucleon cross-section is inversely proportional to the DM mass,
 leading to low sensitivity. In addition, the DM annihilation cross-section is also suppressed by the DM mass and $\la_3$ has to be large in order to 
 reproduce the correct DM  density. 
Thus, for a large DM mass $m_{\text{DM}} \gtrsim 19$ TeV the theory is not perturbative anymore. This roughly agrees with the result of Ref.~\cite{Garcia-Cely:2015khw}, where 22.4 TeV is obtained by considering
all contributions including the gauge bosons.

\section{Conclusions}
\label{sec:conclusion}

In this work, we have explored an interesting correlation between DM, radiative neutrino masses and GUT baryogenesis, based on the scotogenic model~\cite{Ma:2006km}.
The model contains a second Higgs doublet $\Phi$ together with right-handed neutrinos $N$,
both of which are odd under a $Z_2$ symmetry. The lightest one of the $Z_2$-odd particles, $\Phi$, is a DM candidate.
Due to the $Z_2$ symmetry, the type-I seesaw Yukawa coupling of the right-handed neutrinos to the Higgs boson is prohibited but a new coupling $y^\prime \bar{\ell} \Phi N$~($\ell$ is the SM lepton doublet)
is allowed.
Consequently, the neutrino mass is radiatively induced by loops of $\Phi$ and $N$. 
In the context of $(B-L)$-preserving GUT baryogenesis, the additional interaction $\ell \Phi \tol \bar{\ell} \Phi^*$ via
$N$-exchange shifts the $L$ asymmetry into $\Phi$ such that a nonzero $B-L$ asymmetry can be generated.
The net $B-L$ asymmetry will be preserved by the $(B+L)$-violating sphaleron effects and
as a result the observed baryon asymmetry can be obtained.

Moreover, due to the interactions $\Phi^{*} H \tol \Phi H^{*} $ and $\Phi^{(*)}\Phi^{(*)} \tol H^{(*)} H^{(*)} $, the asymmetry in $\Phi$ from $L$ washouts will be further transferred into $H$ that helps to wash out more $L$, leading to a larger $B-L$ asymmetry.
With two Higgs doublets, $\Phi$ and $H$, the induced $B-L$ asymmetry is at most one third~($5/12$ including
the $t$- and $b$-Yukawa coupling effects) of the initial $B+L$ asymmetry from GUT baryogenesis, which is larger than the asymmetry obtained in Ref.~\cite{Fukugita:2002hu},
where the type-I seesaw Yukawa coupling is used to erase $L$ and to produce a nonzero $B-L$ asymmetry. Numerically, we have found that in order to generate a neutrino mass of $\sqrt{ \Delta m^2_{atm}} (= 0.05$ eV)
and achieve $Y^{\text{final}}_{B-L}/Y^{\text{initial}}_{B+L} \gtrsim \mathcal{O}(10^{-2})$, the mass of the right-handed neutrino $M_N$ has to be roughly larger than $10^{13}$ GeV for TeV DM.
If the initial $B+L$ asymmetry is sizable~($\gtrsim 10^{-6}$) and the $B+L$ injection scale is $M_N/10$, $M_N$ can be as low as $10^{10}$ GeV to accommodate both the baryon asymmetry and the neutrino mass, since $L$ washouts can still be efficient enough to create a non-vanishing $B-L$ asymmetry before the sphalerons completely destroy the $B+L$ asymmetry.

We have made sure that with $\Phi$ heavier than 3 TeV one can reproduce the observed relic abundance which requires an $\mathcal{O}(1)$ coupling $\lambda_3$ and at the same time avoid the XENON1T direct search bounds.
In this case $\Phi$ falls out of equilibrium before the electroweak symmetry breaking and the asymmetry stored in $\Phi$ will convert back into $H$. That slightly increases the final baryon asymmetry.
To summarize, we have established an intriguing correlation among GUT baryogenesis, DM phenomenology and neutrino mass mechanism, where the existence of DM revives GUT baryogenesis and
induces the radiative neutrino mass.

\section*{Acknowledgments}
The authors would like to thank Avelino Vicente for pointing out a missing factor of 1/2 in the neutrino mass matrix.
This work is supported by Deutsche Forschungsgemeinschaft (DFG) Grant No. PA 803/10-1.
S.~Zei{\ss}ner was supported by a fellowship of Deutsche Studienstiftung.
W.~C.~Huang is also supported by Danish Council for Independent Research
Grant DFF-6108-00623. The CP3-Origins centre is partially funded by the Danish National Research Foundation, grant number DNRF90.

\bibliography{GUT_nuDM}

\begin{thebibliography}{10}

\bibitem{Georgi:1974sy}
H.~Georgi and S.~L. Glashow.
\newblock {Unity of All Elementary Particle Forces}.
\newblock {\em Phys. Rev. Lett.}, 32:438--441, 1974.

\bibitem{Yoshimura:1978ex}
Motohiko Yoshimura.
\newblock {Unified Gauge Theories and the Baryon Number of the Universe}.
\newblock {\em Phys. Rev. Lett.}, 41:281--284, 1978.
\newblock [Erratum: Phys. Rev. Lett.42,746(1979)].

\bibitem{Toussaint:1978br}
D.~Toussaint, S.~B. Treiman, Frank Wilczek, and A.~Zee.
\newblock {Matter - Antimatter Accounting, Thermodynamics, and Black Hole
  Radiation}.
\newblock {\em Phys. Rev.}, D19:1036--1045, 1979.

\bibitem{Weinberg:1979bt}
Steven Weinberg.
\newblock {Cosmological Production of Baryons}.
\newblock {\em Phys. Rev. Lett.}, 42:850--853, 1979.

\bibitem{Barr:1979ye}
Stephen~M. Barr, Gino Segre, and H.~Arthur Weldon.
\newblock {The Magnitude of the Cosmological Baryon Asymmetry}.
\newblock {\em Phys. Rev.}, D20:2494, 1979.

\bibitem{Klinkhamer:1984di}
Frans~R. Klinkhamer and N.~S. Manton.
\newblock {A Saddle Point Solution in the Weinberg-Salam Theory}.
\newblock {\em Phys. Rev.}, D30:2212, 1984.

\bibitem{Arnold:1987mh}
Peter~Brockway Arnold and Larry~D. McLerran.
\newblock {Sphalerons, Small Fluctuations and Baryon Number Violation in
  Electroweak Theory}.
\newblock {\em Phys. Rev.}, D36:581, 1987.

\bibitem{Arnold:1987zg}
Peter~Brockway Arnold and Larry~D. McLerran.
\newblock {The Sphaleron Strikes Back}.
\newblock {\em Phys. Rev.}, D37:1020, 1988.

\bibitem{Coughlan:1985hh}
G.~D. Coughlan, Graham~G. Ross, R.~Holman, Pierre Ramond, M.~Ruiz-Altaba, and
  J.~W.~F. Valle.
\newblock {Baryogenesis, Proton Decay and Fermion Masses in Supergravity
  {GUTs}}.
\newblock {\em Phys. Lett.}, B158:401--408, 1985.

\bibitem{Babu:1992ia}
K.~S. Babu and R.~N. Mohapatra.
\newblock {Predictive neutrino spectrum in minimal SO(10) grand unification}.
\newblock {\em Phys. Rev. Lett.}, 70:2845--2848, 1993, hep-ph/9209215.

\bibitem{Garbrecht:2005rr}
Bjorn Garbrecht, Tomislav Prokopec, and Michael~G. Schmidt.
\newblock {SO(10)-GUT coherent baryogenesis}.
\newblock {\em Nucl. Phys.}, B736:133--155, 2006, hep-ph/0509190.

\bibitem{Achiman:2007qz}
Yoav Achiman.
\newblock {Spontaneous CP violation in SUSY SO(10)}.
\newblock {\em Phys. Lett.}, B653:325--329, 2007, hep-ph/0703215.

\bibitem{Babu:2012vb}
K.~S. Babu and R.~N. Mohapatra.
\newblock {B-L Violating Nucleon Decay and GUT Scale Baryogenesis in SO(10)}.
\newblock {\em Phys. Rev.}, D86:035018, 2012, 1203.5544.

\bibitem{Babu:2012vc}
K.~S. Babu and R.~N. Mohapatra.
\newblock {Coupling Unification, GUT-Scale Baryogenesis and Neutron-Antineutron
  Oscillation in SO(10)}.
\newblock {\em Phys. Lett.}, B715:328--334, 2012, 1206.5701.

\bibitem{Fukugita:2002hu}
M.~Fukugita and T.~Yanagida.
\newblock {Resurrection of grand unified theory baryogenesis}.
\newblock {\em Phys. Rev. Lett.}, 89:131602, 2002, hep-ph/0203194.

\bibitem{Huang:2016wwj}
Wei-Chih Huang, Heinrich Päs, and Sinan Zeissner.
\newblock {Neutrino assisted GUT baryogenesis - revisited}.
\newblock {\em Phys. Rev.}, D97(5):055040, 2018, 1608.04354.

\bibitem{Ma:2006km}
Ernest Ma.
\newblock {Verifiable radiative seesaw mechanism of neutrino mass and dark
  matter}.
\newblock {\em Phys. Rev.}, D73:077301, 2006, hep-ph/0601225.

\bibitem{Ma:2015xla}
Ernest Ma.
\newblock {Derivation of Dark Matter Parity from Lepton Parity}.
\newblock {\em Phys. Rev. Lett.}, 115(1):011801, 2015, 1502.02200.

\bibitem{Harvey:1990qw}
Jeffrey~A. Harvey and Michael~S. Turner.
\newblock {Cosmological baryon and lepton number in the presence of electroweak
  fermion number violation}.
\newblock {\em Phys. Rev.}, D42:3344--3349, 1990.

\bibitem{Ma:2006fn}
Ernest Ma.
\newblock {Common origin of neutrino mass, dark matter, and baryogenesis}.
\newblock {\em Mod. Phys. Lett.}, A21:1777--1782, 2006, hep-ph/0605180.

\bibitem{Kashiwase:2012xd}
Shoichi Kashiwase and Daijiro Suematsu.
\newblock {Baryon number asymmetry and dark matter in the neutrino mass model
  with an inert doublet}.
\newblock {\em Phys. Rev.}, D86:053001, 2012, 1207.2594.

\bibitem{Kashiwase:2013uy}
Shoichi Kashiwase and Daijiro Suematsu.
\newblock {Leptogenesis and dark matter detection in a TeV scale neutrino mass
  model with inverted mass hierarchy}.
\newblock {\em Eur. Phys. J.}, C73:2484, 2013, 1301.2087.

\bibitem{Racker:2013lua}
J.~Racker.
\newblock {Mass bounds for baryogenesis from particle decays and the inert
  doublet model}.
\newblock {\em JCAP}, 1403:025, 2014, 1308.1840.

\bibitem{Clarke:2015hta}
Jackson~D. Clarke, Robert Foot, and Raymond~R. Volkas.
\newblock {Natural leptogenesis and neutrino masses with two Higgs doublets}.
\newblock {\em Phys. Rev.}, D92(3):033006, 2015, 1505.05744.

\bibitem{Hugle:2018qbw}
Thomas Hugle, Moritz Platscher, and Kai Schmitz.
\newblock {Low-Scale Leptogenesis in the Scotogenic Neutrino Mass Model}.
\newblock 2018, 1804.09660.

\bibitem{Baumholzer:2018sfb}
Sven Baumholzer, Vedran Brdar, and Pedro Schwaller.
\newblock {The New $\nu$MSM : Radiative Neutrino Masses, keV-Scale Dark Matter
  and Viable Leptogenesis with sub-TeV New Physics}.
\newblock 2018, 1806.06864.

\bibitem{Sierra:2013kba}
Diego Aristizabal~Sierra, Chee~Sheng Fong, Enrico Nardi, and Eduardo Peinado.
\newblock {Cloistered Baryogenesis}.
\newblock {\em JCAP}, 1402:013, 2014, 1309.4770.

\bibitem{Griest:1989wd}
Kim Griest and Marc Kamionkowski.
\newblock {Unitarity Limits on the Mass and Radius of Dark Matter Particles}.
\newblock {\em Phys. Rev. Lett.}, 64:615, 1990.

\bibitem{Deshpande:1977rw}
Nilendra~G. Deshpande and Ernest Ma.
\newblock {Pattern of Symmetry Breaking with Two Higgs Doublets}.
\newblock {\em Phys.Rev.}, D18:2574, 1978.

\bibitem{Merle:2015gea}
Alexander Merle and Moritz Platscher.
\newblock {Parity Problem of the Scotogenic Neutrino Model}.
\newblock {\em Phys. Rev.}, D92(9):095002, 2015, 1502.03098.

\bibitem{Vicente:2015zba}
Avelino Vicente.
\newblock {Computer tools in particle physics}.
\newblock 2015, 1507.06349.

\bibitem{Giudice:2003jh}
G.~F. Giudice, A.~Notari, M.~Raidal, A.~Riotto, and A.~Strumia.
\newblock {Towards a complete theory of thermal leptogenesis in the SM and
  MSSM}.
\newblock {\em Nucl. Phys.}, B685:89--149, 2004, hep-ph/0310123.

\bibitem{Moore:2000ara}
Guy~D. Moore.
\newblock {Do we understand the sphaleron rate?}
\newblock In {\em {Strong and electroweak matter. Proceedings, Meeting, SEWM
  2000, Marseille, France, June 13-17, 2000}}, pages 82--94, 2000,
  hep-ph/0009161.

\bibitem{D'Onofrio:2014kta}
Michela D'Onofrio, Kari Rummukainen, and Anders Tranberg.
\newblock {Sphaleron Rate in the Minimal Standard Model}.
\newblock {\em Phys. Rev. Lett.}, 113(14):141602, 2014, 1404.3565.

\bibitem{Khlebnikov:1988sr}
S.~{\relax Yu}. Khlebnikov and M.~E. Shaposhnikov.
\newblock {The Statistical Theory of Anomalous Fermion Number Nonconservation}.
\newblock {\em Nucl. Phys.}, B308:885--912, 1988.

\bibitem{Ade:2015xua}
P.~A.~R. Ade et~al.
\newblock {Planck 2015 results. XIII. Cosmological parameters}.
\newblock {\em Astron. Astrophys.}, 594:A13, 2016, 1502.01589.

\bibitem{Buchmuller:2005eh}
W.~Buchmuller, R.~D. Peccei, and T.~Yanagida.
\newblock {Leptogenesis as the origin of matter}.
\newblock {\em Ann. Rev. Nucl. Part. Sci.}, 55:311--355, 2005, hep-ph/0502169.

\bibitem{Arhrib:2013ela}
Abdesslam Arhrib, Yue-Lin~Sming Tsai, Qiang Yuan, and Tzu-Chiang Yuan.
\newblock {An Updated Analysis of Inert Higgs Doublet Model in light of the
  Recent Results from LUX, PLANCK, AMS-02 and LHC}.
\newblock {\em JCAP}, 1406:030, 2014, 1310.0358.

\bibitem{Ilnicka:2015jba}
Agnieszka Ilnicka, Maria Krawczyk, and Tania Robens.
\newblock {Inert Doublet Model in light of LHC Run I and astrophysical data}.
\newblock {\em Phys. Rev.}, D93(5):055026, 2016, 1508.01671.

\bibitem{Khan:2015ipa}
Najimuddin Khan and Subhendu Rakshit.
\newblock {Constraints on inert dark matter from the metastability of the
  electroweak vacuum}.
\newblock {\em Phys. Rev.}, D92:055006, 2015, 1503.03085.

\bibitem{Belyaev:2016lok}
Alexander Belyaev, Giacomo Cacciapaglia, Igor~P. Ivanov, Felipe Rojas-Abatte,
  and Marc Thomas.
\newblock {Anatomy of the Inert Two Higgs Doublet Model in the light of the LHC
  and non-LHC Dark Matter Searches}.
\newblock {\em Phys. Rev.}, D97(3):035011, 2018, 1612.00511.

\bibitem{Eiteneuer:2017hoh}
Benedikt Eiteneuer, Andreas Goudelis, and Jan Heisig.
\newblock {The inert doublet model in the light of Fermi-LAT gamma-ray data: a
  global fit analysis}.
\newblock {\em Eur. Phys. J.}, C77(9):624, 2017, 1705.01458.

\bibitem{Garcia-Cely:2015khw}
Camilo Garcia-Cely, Michael Gustafsson, and Alejandro Ibarra.
\newblock {Probing the Inert Doublet Dark Matter Model with Cherenkov
  Telescopes}.
\newblock {\em JCAP}, 1602(02):043, 2016, 1512.02801.

\bibitem{Jungman:1995df}
Gerard Jungman, Marc Kamionkowski, and Kim Griest.
\newblock {Supersymmetric dark matter}.
\newblock {\em Phys. Rept.}, 267:195--373, 1996, hep-ph/9506380.

\bibitem{Bertone:2004pz}
Gianfranco Bertone, Dan Hooper, and Joseph Silk.
\newblock {Particle dark matter: Evidence, candidates and constraints}.
\newblock {\em Phys. Rept.}, 405:279--390, 2005, hep-ph/0404175.

\bibitem{Aprile:2017iyp}
E.~Aprile et~al.
\newblock {First Dark Matter Search Results from the XENON1T Experiment}.
\newblock {\em Phys. Rev. Lett.}, 119(18):181301, 2017, 1705.06655.

\bibitem{Cline:2013gha}
James~M. Cline, Kimmo Kainulainen, Pat Scott, and Christoph Weniger.
\newblock {Update on scalar singlet dark matter}.
\newblock {\em Phys. Rev.}, D88:055025, 2013, 1306.4710.
\newblock [Erratum: Phys. Rev.D92,no.3,039906(2015)].

\bibitem{Cui:2017nnn}
Xiangyi Cui et~al.
\newblock {Dark Matter Results From 54-Ton-Day Exposure of PandaX-II
  Experiment}.
\newblock {\em Phys. Rev. Lett.}, 119(18):181302, 2017, 1708.06917.

\bibitem{Akerib:2016vxi}
D.~S. Akerib et~al.
\newblock {Results from a search for dark matter in the complete LUX exposure}.
\newblock {\em Phys. Rev. Lett.}, 118(2):021303, 2017, 1608.07648.

\bibitem{Aprile:2018dbl}
E.~Aprile et~al.
\newblock {Dark Matter Search Results from a One Tonne$\times$Year Exposure of
  XENON1T}.
\newblock 2018, 1805.12562.

\end{thebibliography}
\bibliographystyle{hunsrt}

\end{document}